\documentclass[a4paper,11pt]{article}

\usepackage[square,numbers,merge,comma,sort&compress]{natbib} 
\usepackage{pos}
\usepackage{mathrsfs,slashed}
\DeclareMathAlphabet{\mathpzc}{OT1}{pzc}{m}{it}
\DeclareMathAlphabet{\stmathcal}{OMS}{cmsy}{m}{n}

\title{Supersymmetric theories and graphene}

\author*[a,b]{Antonio Gallerati\,}

\affiliation[a]{Politecnico di Torino, Dipartimento di Scienza Applicata e Tecnologia,\\
corso Duca degli Abruzzi 24, 10129 Torino, Italy}

\affiliation[b]{Istituto Nazionale di Fisica Nucleare, Sezione di Torino,\\
via Pietro Giuria 1, 10125 Torino, Italy}

\emailAdd{antonio.gallerati@polito.it}

\abstract{We discuss a 1+2 dimensional model with unconventional supersymmetry at the boundary of an AdS${}_4$, \,$\mathcal{N}$-extended supergravity. The resulting features of the supersymmetric boundary open the possibility of describing the electronic properties of graphene-like 2D materials at the Dirac points \textbf{K} and \textbf{K'}, exploiting a top-down approach. The Semenoff and Haldane-type masses entering the corresponding Dirac equations can be then extrapolated from the geometric parameters of the model describing the substrate.}

\FullConference{%
  40th International Conference on High Energy physics - ICHEP2020\\
  July 28 - August 6, 2020\\
  Prague, Czech Republic (virtual meeting)
}

\begin{document}

\maketitle

\section{Introduction}
The recent developments in the studies of the physics of bidimensional materials like graphene provide a new connection between condensed matter and theoretical quantum field theories, opening a window on the possibility of a direct observation of quantum behaviour in the curved background of a solid state system \cite{katsnelson2007graphene,geim2007rise,CastroNeto:2009zz}.
In 1984 Semenoff formulated the hypothesis that graphene could realize the physics of two dimensional massless Dirac fermions \cite{Semenoff:1984dq}, this property discriminating graphene from other 2D system. General, graphene-like bidimensional materials turn out to realize the physics of spinorial fields, whose Dirac properties emerge due to the structure of the honeycomb lattice with which the charge carriers interact, determining a natural description of its electronic properties in terms of Dirac pseudoparticles living in a 1+2 dimensional curved spacetime background
\cite{Novoselov:2005kj,CastroNeto:2009zz,Cortijo:2006mh,Gallerati:2018dgm,Gallerati:2021rtp}.
\par\smallskip
In the following, we will consider a Chern-Simons theory in 1+2 dimensions, that turns out to be an effective theory for a Dirac spin-1/2 fermion, defined on a curved geometry. The fermion is minimally coupled to the background gravity and the system features an unconventional form of supersymmetry based on a graded Lie algebra, where a peculiar ansatz for the fermionic $D=3$ gauge fields $\psi^A$ is assumed \cite{Alvarez:2011gd,Andrianopoli:2019sip}, expressing them as composite fields of the vielbein $e_i$ and spin 1/2 fields $\chi^A$,
\begingroup
\setlength{\abovedisplayshortskip}{-4pt plus 1pt}%
\setlength{\abovedisplayskip}{-4pt plus 1pt minus 4pt}
\setlength{\belowdisplayshortskip}{8pt plus 1pt}%
\setlength{\belowdisplayskip}{8pt plus 1pt minus 1pt}%
\begin{equation}
\psi^A=i\,\gamma_i\,e^i\chi^A.
\label{eq:AVZ}
\end{equation}
\endgroup
%
The Lagrangian associated with the model is a Chern-Simons form and the resulting model is particularly suited for describing graphene near the Dirac points in a generic lattice with nonvanishing curvature and torsion \cite{Andrianopoli:2019sip}.
In particular, the chosen three-dimensional model of unconventional supersymmetry is defined at the boundary of an AdS${}_4$ supergravity vacuum, and could be holographically realized as the boundary theory of an $\mathcal{N}$-extended four-dimensional supergravity of the AdS${}_4$ spacetime.
Applying the above ansatz \eqref{eq:AVZ} for the fermionic gauge fields, an effective model for the massive spin-1/2 fields on a curved background is obtained: this allows to introduce extra internal degrees of freedom which can provide an application of the model to the description of graphene, where the Semenoff and Haldane microscopic models can be realized.

\section{Microscopic models}\label{sec:micro}
A graphene sheet consists in a bidimensional system made of carbon atoms arranged in a honeycomb lattice.
The bidimensional hexagonal lattice is split into two inequivalent, interpenetrating triangular sublattices, so that the unit cell is made of two adjacent atoms belonging to different sublattices.
The first Brillouin zone (FBZ) of the reciprocal lattice
has the same hexagonal form of the honeycomb lattice and is rotated by a $\pi/2$ angle. The corners of the FBZ can be divided into two inequivalent classes, since only two of the six can be chosen to be independent, the remaining four being connected to them by a reciprocal lattice vector; this means we can consider only two inequivalent corners, referred to as ``valleys'', labeled $\mathbf{K}$ and $\mathbf{K}'$.

\paragraph{Pristine graphene.}
Pure graphene quantum states can be formulated in terms of the \emph{tight-binding model}, describing electrons hopping in the (single-state per site) honeycomb lattice. The formulation describes the limit of far apart ions, the single-particle eigenstates referring to electrons affected by a single corresponding ion.
In this picture, electrons can only tunnel to their first neighbor atoms, with a hopping amplitude $t_1$
and the system is characterized by the tight-binding Hamiltonian:
\begin{equation}
{H}_1=t_1\,\sum_{\langle i,j\rangle}\, c_i^\dagger\,c_j\;,
\end{equation}
where the creation (annihilation) operator $c_i^\dagger=c^\dagger(\mathbf{r}_i)$ \,\big($c_i=c(\mathbf{r}_i)$\big) acts on site $\mathbf{r}_i$, while the sum
$\langle i,j\rangle$ only runs on nearest neighbors sites $\mathbf{r}_i$, $\mathbf{r}_j$.

\subsection{Mass gaps}\label{subsec:massgaps}
Mass terms for two-dimensional samples can be induced in several ways, in particular when some symmetries of the system are broken.
In particular, mass gaps at the Dirac points for two-dimensional graphene-like systems can be obtained from generalizations of the tight binding microscopic model. This gap generation was first discussed by Semenoff, introducing a mass term through an on-site dependent potential $\pm M$, spoiling sublattices equivalence and  breaking then parity symmetry of the theory \cite{Semenoff:1984dq}.
Another microscopic model was proposed by Haldane, including local magnetic fields over the honeycomb hexagon, breaking time-reversal symmetry of the model \cite{Haldane:1988zza}.

\paragraph{Haldane model.}
The formulation of the Haldane model was motivated by the realization of a quantum anomalous Hall effect
and is achieved by the introduction of periodic local magnetic flux densities, with zero net flux over the hexagon lattice cell.
The microscopic Hamiltonian is modified with second-neighbor hopping terms with unimodular (chiral) phase factors
and has the form
\begin{equation}
H \,=\, H_1+H_2 \,=\, H_1 \,+\, t_2\!\!\!\sum_{\;\;\langle i,j\rangle_{{}_{(2)}}}
\!e^{i\,\varphi\,\alpha_{ij}}\,c_i^\dagger\,c_j \,+\, \epsilon_i\,M\,\sum_i\,c_i^\dagger\,c_i\:,
\qquad \epsilon_i=\pm1\:,
\label{HamHald}
\end{equation}
where $H_1$ is the tight binding Hamiltonian and $H_2$ accounts for the local magnetic fields (breaking time reversal symmetry), and for a Semenoff parity-breaking term. The first sum in $H_2$ runs on second nearest neighbors sites, $t_2$ being the hopping amplitude, while the second term is the Semenoff contribution coming from on-site potential energy $M$, the prefactor $\epsilon_i=\pm1$ depending on whether the site $i$ is on the first or second sublattice.
The fermion masses in the two inequivalent valleys \textbf{K}, \textbf{K'} turn out to be \cite{Haldane:1988zza}
\begingroup
\setlength{\abovedisplayshortskip}{2pt plus 1pt}%
\setlength{\abovedisplayskip}{2pt plus 1pt minus 4pt}
\begin{equation}
m_{{}_{\textbf{K},\textbf{K'}}}=M \mp 3\sqrt{3}\:t_2 \sin{\varphi}\;.
\end{equation}
\endgroup

\section{Geometrical top-down approach}
Our interest here focuses on the AVZ model \cite{Alvarez:2011gd}, a Chern-Simons system in $1+2$ dimensions featuring interesting properties \cite{Achucarro:1987vz}, particularly in connection with the holographic correspondence. 
The model acts as an effective theory for a massive spin-1/2 fermion, generically defined on a curved geometry and minimally coupled to the background gravity, particularly suited for describing graphene near the Dirac points in a generic spatial lattice with non-vanishing curvature and torsion.
In particular, the system exhibits an unconventional form of supersymmetry based on a graded Lie algebra,
whose $+$ and $-$ sectors can be naturally interpreted as related to the graphene Dirac points \textbf{K}, \textbf{K'}.

\paragraph{Holographic formulation.}
The Chern-Simons theory of the AVZ model, originally constructed for the $\mathcal{N}=2$ case, can be extended to a $D=3$ model with $\mathcal{N}$ supersymmetries on AdS${}_3$ \cite{Andrianopoli:2019sip}. The three-dimensional model can be in turn holographically realized as the boundary theory of a $D=4$ supergravity in AdS$_4$, exploiting suitable choices for the boundary conditions of the four-dimensional fields \cite{Andrianopoli:2019sip}. Imposing the AVZ ansatz \eqref{eq:AVZ} for the $D=3$ fermions identifies the resulting spin-$1/2$ fields $\chi$ as the radial component of the four-dimensional gravitini, whose mass is related to the AdS$_3$ radius. Applying then the resulting model to the effective description of the electronic properties of graphene-like 2D materials, provides a top-down approach to the understanding of the (supersymmetric) origin of the physical system phenomenology, in that the effective $D=3$ theory, derived at the boundary and defining a Dirac fermion living in 1+2 dimensions, originates from a well-defined effective supergravity in the bulk.

\subsection{The model}\label{subsec:model}
The starting point is an AdS$_4$ vacuum of an $\mathcal{N}$-extended pure supergravity theory preserving all $\mathcal{N}$ supersymmetries. The vacuum symmetry is described by the supergroup $\mathrm{OSp}(\mathcal{N}|4)$ group and we require all scalar and spin-$1/2$ fields at the conformal boundary to be frozen at their vacuum values, and that the remaining fields obey the $\mathfrak{osp}(\mathcal{N}|4)$ Maurer-Cartan equations.
The dual description of the $\mathfrak{osp}(\mathcal{N}|4)$ algebra is given in terms of the connection
\begingroup
\setlength{\abovedisplayshortskip}{8pt plus 1pt}%
\setlength{\abovedisplayskip}{8pt plus 1pt minus 1pt}
\setlength{\belowdisplayshortskip}{9pt plus 1pt}%
\setlength{\belowdisplayskip}{9pt plus 1pt minus 1pt}%
\begin{equation}
\mathbb{A}=\theta^i\otimes\mathpzc{E}_i
    =\frac{1}{2}\,\omega^{\mathcal{A}\mathcal{B}}\,\stmathcal{L}_{\mathcal{A}\mathcal{B}}+
\frac{1}{2}\,A^{CD}\,T_{CD}+\bar\Psi_{\alpha}^A\,Q_A^{\alpha},
\end{equation}
\endgroup
where $\stmathcal{L}_{\mathcal{A}\mathcal{B}}$ ($\mathcal{A},\mathcal{B}=0,\dots,4$) and $T_{CD}$ ($C,D=1,\dots,\mathcal{N}$) are the $\mathrm{SO}(2,3)$ and $\mathrm{SO}(\mathcal{N})$ generators, while $\omega^{\mathcal{A}\mathcal{B}}$, $A^{CD}$ and $\Psi^A$ are one-forms, and $Q_A^{\alpha}$ ($\alpha=1,\dots,4$) are the Majorana supersymmetry generators. The structure of the algebra is encoded in the Maurer-Cartan equations $d\mathbb{A}+\mathbb{A}\wedge \mathbb{A}=0$.\par
If we decompose the AdS$_4$ superalgebra in terms of $\mathrm{SO}(1,1)\times \mathrm{SO}(1,2)\subset\mathrm{SO}(2,3)$, we can express the asymptotic algebra in a way in which the $\mathrm{SO}(1,1)$ grading of the fields is manifest. In particular, we write the $D=4$ vielbein and $\Psi^A$ form expansions at the AdS$_4$ boundary ($r\rightarrow\infty$) as
\begingroup
\setlength{\abovedisplayshortskip}{5pt plus 1pt}%
\setlength{\abovedisplayskip}{6pt plus 1pt minus 1pt}
\setlength{\belowdisplayshortskip}{5pt plus 1pt}%
\setlength{\belowdisplayskip}{6pt plus 1pt minus 1pt}%
\begin{equation}
E^i_{\pm}\simeq\pm\frac12\left(\frac{r}{\ell}\right)^{\pm1}E^i,
\quad\;\;
E^i= f\:e^i\,;
\qquad
\Psi^A=\Psi^A_{+}+\Psi^A_{-}\,,
\quad\;
\Psi^A_{\pm\,\mu}\,dx^\mu\simeq\frac{1}{\sqrt{2}}\left(\frac{r}{\ell}\right)^{\pm\frac{1}{2}}\psi^A,
\quad
\end{equation}
\endgroup
expressed in the basis of the $D=3$ vielbein $e^i$, $f$ being an indeterminate function and $\ell$ the AdS radius. It is then possible to show that the Maurer-Cartan equations at the boundary describe the superalgebra of $\mathrm{OSp}(p|2)_{+}\times\mathrm{OSp}(q|2)_{-}$ with $p+q=\mathcal{N}$ \cite{Andrianopoli:2019sip}, the resulting $D=3$ world volume describing a generalized AVZ model featuring a local Nieh-Yan-Weyl symmetry (NYW local scale invariance \cite{Nieh:1981xk}).
In particular, we can write for the 1+2 dimensional torsion the expression \cite{Andrianopoli:2019sip}
\begin{equation}
T^i_{\pm}=\mathcal{D}[\Omega_\pm]\,e^i
    =\beta\,e^i+\tau_\pm\,\epsilon^{ijk}\,e_j\wedge e_k\,,
\label{eq:tor}
\end{equation}
where $\beta$ and $\tau_\pm$ are 1 and 0-forms, respectively, having defined the world-volume gauge field
\,$\Omega^i_\pm=\frac{1}{2}\,\epsilon^{ijk}\,\omega_{jk}\pm E^i/\ell$\, in terms of the boundary value $\omega_{jk}$ of the spin connection.
One can also compute the covariant derivative of $e^i$ with respect to $\omega^i:=\frac{1}{2}\,\epsilon^{ijk}\,\omega_{jk}$, %
\begin{equation}
\mathcal{D}[\omega]\,e^i=\beta\,e^i+\tau\,\epsilon^{ijk}\,e_j \wedge e_k\:,
\end{equation}
whose parameter $\tau$ can be shown to be related to that of the above \eqref{eq:tor} as \,${\tau_{+} + 2\,\tfrac{f}{\ell}=\tau_{-} - 2\,\tfrac{f}{\ell} \equiv \tau}$.
The existing NYW symmetry can be used to set $\beta=0$ locally on any open neighborhood of the $D=3$ world volume, imposing a non-trivial condition on the topology of spacetime.\par
%
The algebra of the ${\mathrm{OSp}(p|2)_{+}\times\mathrm{OSp}(q|2)_{-}}$ supergroup implies the relation ${\mathcal{D}[\Omega_\pm,\,A_\pm]\psi_\pm=0}$, having defined the covariant derivative
\begin{equation}
{\mathcal{D}[\Omega_{\pm},\,A_{\pm}]\,\psi_\pm:=d\psi_{\pm}+\frac{i}{2}\,\Omega^i_{\pm}\,\gamma_i\,\psi_{\pm}+A_{\pm}\,\psi_{\pm}}\,,
\end{equation}
and having denoted \,${\psi_{+}:=\psi^{a_1}}$, \,${\psi_{-}:=\psi^{a_2}}$, \,${A_{+}:=A^{a_1 b_1}}$, \,${A_{-}:=A^{a_2 b_2}}$, with \,${a_1,b_1=1,\dots,p}$, \,${a_2,b_2=p+1,\dots,\mathcal{N}}$, \,reflecting the structure of the $\mathfrak{osp}(p|2)_{+}\oplus\mathfrak{osp}(q|2)_{-}$ superalgebra.
If we consider the AVZ ansatz \eqref{eq:AVZ}, we obtain for the world volume spinorial fields $\chi_A=(\chi_{a_1},\chi_{a_2})$ the Dirac equation \cite{Andrianopoli:2019sip}
\begin{equation}
\slashed{\mathcal{D}}[\omega^\prime,\,A_\pm]\,\chi_\pm=-\frac{3}{2}\,i\,\tau_\pm\,\chi_\pm\,,
\label{eq:Dirac}
\end{equation}
\sloppy
the covariant derivative being expressed in terms of the torsion-free Lorentz connection ${\omega^{\prime i}=\Omega_{\pm}^i+\tau_{\pm}\,e^i}$, having denoted \,${\chi_{+}=\chi^{a_1}}$, \,${\chi_{-}=\chi^{a_2}}$.

\paragraph{\boldmath Special case $p=q$. \unboldmath}
Let us consider the case of even integers $p$ and $q$, since this allows to arrange the real spinors $\chi_{\pm}$ into $p/2$ and $q/2$ Dirac spinors.
If we restrict to the $p=q$ case, parity (spatial reflection in $D=3$) becomes a symmetry of the model, leaving the field equations invariant. The $+$ and $-$ sectors turn out to be related by a reflection symmetry in one spatial axis and, in the context of graphene-like materials, can be naturally associated with the inequivalent corners of the FBZ, that is we interpret them as referring to the $\textbf{K}$, $\textbf{K'}$ inequivalent valleys.
In particular, we can also notice that a parity transformations in pure graphene exchanges unit cell sites in the honeycomb lattice, mapping also the two valleys into each other if the reflection is combined with a time-reversal transformation. In bidimensional materials with inequivalent honeycomb sites, parity reflection is not a symmetry of the system, implying the existence of (at least) a parity-violating Semenoff mass term in the effective Dirac equation.

\subsection{Results}
From the perspective of the description of general, 1+2 dimensional graphene-like systems, the introduced spin 1/2 fields $\chi_\pm$ satisfying the Dirac equations \eqref{eq:Dirac} can be used to suitably characterize the sample long-wavelength charge carriers. In particular, the fermion effective masses turn out to be
\begin{equation}
m_{\pm}=\frac{3}{2}\,\tau_{\pm}=\frac{3}{2}\left(\tau\mp2\,\frac{f}{\ell}\right)\,,
\end{equation}
then expressed in terms of the torsion and geometric parameters of the model.
In light of the discussion of the previous section \ref{subsec:model}, the $p=q$ choice allows the identification of the $\pm$ sectors with the $\textbf{K}$, $\textbf{K'}$ valleys in graphene-like materials,
\begin{equation}
m_{\pm}=m_{{}_{\textbf{K},\textbf{K'}}}=M \mp 3\sqrt{3}\:t_2 \sin{\varphi}\;.
\end{equation}
This also leads to the identification of the parity-even and odd components of the corresponding masses with Semenoff and Haldane-type mass contributions, respectively (see \ref{subsec:massgaps}), implying in particular the relations
\begin{equation}
M\rightarrow\,\frac{3}{2}\,\tau\,,\qquad\;
\sqrt{3}\:t_2\,\sin(\varphi)\rightarrow\,\frac{f}{\ell}\,.
\end{equation}
Summarizing, we have seen how the extension of unconventional supersymmetry to the $\mathrm{OSp}(p|2)\times\mathrm{OSp}(p|2)$ superalgebra can be instrumental in describing the electronic properties of graphene-like systems in the inequivalent valleys, connecting them to physical situations in which the symmetry between them is broken. This can happen, for instance, if a substrate with inequivalent sites and/or suitable local magnetic fields are present, so that parity and/or time reversal symmetry are broken, giving then rise to Semenoff and Haldane-type mass terms. One of the main results we have presented is the embedding of this effective description in an $\mathcal{N}$-extended, $D=4$ supergravity, setting the stage for a more detailed holographic analysis which will be pursued in a future work.

\bibliographystyle{JHEP}
\bibliography{bibliografia}

\end{document}